\documentclass[%
 reprint,
 amsmath,amssymb,
 aps,
longbibliography
]{revtex4-2}
\pdfoutput=1
\usepackage{graphicx}
\usepackage{dcolumn}
\usepackage{bm}
\usepackage{slashed}
\usepackage{hyperref}

\def\be{\begin{eqnarray}}
\def\ee{\end{eqnarray}}

\def\Vol{\operatorname{Vol}}

\def\d{{\rm d}}

\def\abs#1{{\left| #1 \right| }}
\def\expval#1{{\left\langle #1 \right\rangle }}
\def\pdv#1#2{{\frac{\partial #1}{\partial #2}}}

\begin{document}


\title{The Tilting Space of Boundary Conformal Field Theories}

\author{Christopher P.\ Herzog}
\author{Vladimir Schaub}%
\affiliation{%
Mathematics Department, King's College London, The Strand WC2R 2LS, England
}%

\date{\today}

\begin{abstract}
In boundary conformal field theories, global symmetries can be broken by boundary conditions, generating a homogeneous conformal manifold. We investigate these geometries, showing they have a coset structure, and give fully worked out examples in the case of free fields of spin zero and one-half. These results give a simple illustration of the salient features of conformal manifolds while generalising to interacting and more intricate setups. 
Our work was inspired by \cite{Drukker:2022pxk}.
\end{abstract}

\maketitle

Conformal field theories (CFTs) are useful for describing critical phenomena, fixed points of the renormalization group in
quantum field theory more generally, and potentially even quantum gravity.  
Occasionally, a CFT can appear as a member of a family.
This situation occurs when 
the theory possesses one or more marginal couplings, i.e.\ couplings that do not scale under the action of 
the dilation generator of the conformal group.  
Two well known examples of such marginal couplings 
are the compactification radius of a compact scalar in two dimensions 
and the gauge coupling of maximally supersymmetric Yang-Mills theory in four dimensions.
These couplings can be interpreted as coordinates on a manifold, which in turn 
puts interesting constraints on the correlation functions of such theories
\cite{Seiberg:1988pf,Kutasov:1988xb,Ranganathan:1992nb}.
%
%
Finding exactly marginal parameters for CFTs that are neither supersymmetric nor two dimensional has historically been a challenge.
%
Recently, however, it was pointed out \cite{Drukker:2022pxk} that defect and boundary conformal field theories with global symmetry, where the defect or boundary breaks the global symmetry, naturally have such couplings.
The role of conserved currents in providing exactly marginal operators has long been known in boundary two dimensional CFT and supersymmetric
CFTs  in higher dimensions \cite{Recknagel:1998ih,Green:2010da}.  It should perhaps come as no surprise then that these currents play also a role for 
defect and boundary CFTs in higher dimensions as well.  

In contrast to ref.\  \cite{Drukker:2022pxk} where the focus was on line and surface defects, 
the goal of this letter is to investigate the nature of these couplings in boundary conformal field theories (bCFTs).
%
%
 We focus on two 
particularly simple examples: $N$ free massless scalar or spinor fields where the coupling acts by changing the boundary 
conditions.  
By looking at free theories, we will be able to investigate the correlation functions not just perturbatively in the value of the marginal couplings but exactly, for any value.  As a result, we will be able to write down completely
explicit expressions for the metric, connection, and curvature of the conformal manifold \footnote{%
For earlier work on marginal operators in $d>2$ boundary and defect CFT, see 
A.~Karch and Y.~Sato, 
\href {\doibase 10.1007/JHEP07(2018)156}{JHEP {\bf 07}, 156 (2018)}, 
\href{http://arxiv.org/abs/1805.10427}{arXiv:1805.10427 [hep-th]}.
%
}.

Here is an instance of the more general insight of ref.\ \cite{Drukker:2022pxk} for the particular case of a 
$U(1)$ global symmetry.  In the absence of a defect, 
Noether's Theorem implies the existence of a corresponding conserved current
$J^\mu(x)$.   The defect is assumed to break this symmetry
\be
\partial_\mu J^\mu(x) = T(x_\parallel) \delta(x_\perp) \ ,
\ee
where $T(x_\parallel)$ is the `tilt' operator  living on the defect 
 and $\delta(x_\perp)$ is a Dirac delta function localized on 
the defect.  (In the case of a boundary, $T(x_\parallel)$ can be traded for the boundary value of normal component of the current,
$J^\perp(x_\parallel)$, by Gauss's Law.)  The presence of a defect cannot change the scaling dimension of the otherwise conserved current
 $J^\mu(x)$, which thus must be $d-1$ in $d$ dimensions.  By dimensional analysis $T(x_\parallel)$ will have scaling dimenion $n$ for a $n$-dimensional defect.  Given the existence of $T(x_\parallel)$, we can deform the path integral for the original theory by inserting the tilt:
 \be
 \exp \lambda \int \d^{n} x_\parallel \, T(x_\parallel) 
 \ee
 where $\lambda$ is a marginal coupling.  
 
Remarkably, the associated conformal manifold in this case 
can be absorbed by a field redefinition.  
The situation is similar to spontaneous symmetry breaking, where even though there is a moduli space of vacua,
they can be related to each other through the action of the broken generators of the group.
Indeed, let us couple the current to an external gauge potential $A_\mu(x)$:
\be
\int \d^d x \, A_\mu(x) J^\mu (x) \ .
\ee
Under a gauge transformation $A_\mu \to A_\mu -  \partial_\mu \lambda$, this coupling changes by the defect localized term
\be
\int \d^{n} x_\parallel \, T(x_\parallel) \lambda(x_\parallel) \ .
\ee
Constant gauge transformations $\lambda(x_\parallel)  = \lambda$ correspond to global symmetry transformations under the $U(1)$ symmetry.  Thus, we anticipate that this coupling can be removed by redefining the bulk fields of the CFT under a $U(1)$ transformation.  Indeed, we will be able to see this redefinition explicitly for the free scalar and spinor bCFTs below.

Before engaging in the details, it is worth mentioning that bCFT offers an even simpler way of seeing that
the deformation $\lambda \int \d^{d-1} x \, T(x_\parallel)  = \lambda \int \d^{d-1}x \, J^\perp (x_\parallel)$ can be removed by a field redefinition.  In this case, the operator
\be
\exp \lambda \int \d^{d-1} x_\parallel \, J^\perp(x_\parallel) 
\ee
can be interpreted as an element of the $U(1)$ group.   The integral $\int \d^{d-1} x \, J^\perp(x_\parallel)$ is the conserved charge associated with the boundary foliation of the space-time, and hence an element of the Lie algebra.  The exponential is then the usual exponential map for Lie groups that converts a Lie algebra element into a group element, acting on the rest of the theory.  Clearly, it can be ``undone'' by acting with its inverse.
The generalization of this argument to nonabelian Lie groups is straightforward, and we will not belabor it here.

Despite the almost trivial nature of the conformal manifold in these defect and boundary cases, 
it nevertheless exemplifies the properties of conformal manifolds more generally. That one can perform a field redefinition to reabsorb the coupling means that the ensuing manifold is homogeneous. The two point function of the marginal tilt operators furnishes a metric on the manifold \cite{Zamolodchikov:1986gt}.  The three point functions vanish up to contact terms but are associated with a connection on said manifold \cite{Seiberg:1988pf,Kutasov:1988xb}. Finally, the four point function can be used to compute the Riemann curvature \cite{Kutasov:1988xb}.
In what follows, we first analyze the case of free scalar with a boundary and then, more briefly, the case of a free spinor. We end with an argument that the manifolds we compute should be largely insensitive to the addition of interactions and propose some further generalisations.

\vskip 0.1in
\noindent
{\it Scalar Moduli}

The starting point is a system of $N$ free scalar fields $\phi_I$, $I = 1, \ldots, N$ in the presence of a boundary along $x^\perp = 0$.  To $p$ of the scalars $\phi_a$, $a = 1, \ldots, p$, we apply Dirichlet boundary conditions.  To $q = N-p$ of the 
scalars $\phi_i$, $i = p+1, \ldots, N$, we apply Neumann boundary conditions.  
In this case, the presence of the boundary has broken the $O(N)$ global symmetry group to $O(p) \times O(q)$ through
boundary conditions.
%
%
There are then a collection of marginal operators $\phi_i \partial_\perp \phi_a$ and associated marginal
couplings $\lambda_{ai}$ that we can use to ``rotate'' the boundary conditions.  

Given the global $O(N)$ symmetry, there are current operators $J^\mu_{IJ} = \phi_I \partial^\mu \phi_J - \phi_J \partial^\mu \phi_I$.  The marginal operators $\phi_i \partial_\perp \phi_a$ can then be understood as the restriction of $J^\perp_{ia}$ to the boundary $x^\perp = 0$, where $\phi_a$ and $\partial_\perp \phi_i$ vanish by the boundary conditions.  

A useful string theory inspired picture to keep in mind is a fundamental string ending on a D-string.  In this case, the bCFT is the two-dimensional worldsheet of the string, and the fields $\phi_I$ are embedding coordinates of the string in $N$ dimensional spacetime.  By rotating the D-string ninety degrees, we convert a Neumann boundary condition into a Dirichlet one and vice-versa. 
%
%

To be more specific, our system is described by the following classical action
\be
\label{Sscalar}
S = S_{\rm bulk} + S_{\rm bry}
\ee
where
\be
S_{\rm bulk} = \frac{1}{2} \int_{x^\perp>0} \d^d x \, (\partial_\mu \phi) \cdot (\partial^\mu \phi) \ ,
\ee
and
\be
S_{\rm bry}  =  \int_{x^\perp=0} \d^{d-1} x \, \left( \phi_a  \partial_\perp \phi_a + \lambda_{ai} \phi_i \partial_\perp \phi_a \right) \ ,
\ee
where a sum on repeated indices is implied.  
The $\phi_a \partial_\perp \phi_a$ term is added to impose Dirichlet conditions on the $\phi_a$ scalars in the absence of the $\lambda_{ai}$ couplings.
More generally, the variational principle imposes  the boundary conditions
\be
\label{scalarbc}
\begin{split}
\lambda_{ai} \partial_\perp \phi_a - \partial_\perp \phi_i &=& 0 \ , \\
\lambda_{ai} \phi_i + \phi_a &=& 0 \ .
\end{split}
\ee
We impose no condition on the variations $\delta \phi$ and $\partial_n \delta \phi$, and thus interpret their
coefficients as boundary equations of motion to derive the above equations.

To see explicitly that the $\lambda_{ai}$ can be removed by a field redefinition, 
 it is convenient to write the boundary term using the matrix
\be
A\equiv \left(
\begin{array}{cc}
{\rm id}_p & 0 \\
\lambda^T & 0
\end{array}
\right)
\ee  
such that $S_{\rm bry}  =  \int_{x^\perp=0} \d^{d-1} x \, \phi \cdot A \cdot \partial_\perp \phi$. 
We use the notation that ${\rm id}_p$ is a $p \times p$ identity matrix.

Defining 
\be
C \equiv  \left( \begin{array}{cc}
-\frac{1}{1 + \lambda \lambda^T} \lambda \lambda^T &\frac{1}{1 + \lambda \lambda^T}  \lambda \\
-\frac{1}{1 + \lambda^T \lambda} \lambda^T \lambda \lambda^T & \frac{1}{1 + \lambda^T \lambda} \lambda^T \lambda
\end{array}
\right) \ ,
\ee
and
given the boundary conditions (\ref{scalarbc}), we are free to add the redundant operator
$\int \d^{d-1} x \, \phi \cdot C \cdot   \partial_\perp \phi$
 to the action without affecting correlation functions or boundary conditions.
The identity $\lambda (1 + \lambda^T \lambda)^{-1} = (1+\lambda \lambda^T)^{-1} \lambda$ is useful in checking this claim.

The effect of adding $C$ is to symmetrize the matrix $A$:
\be
\begin{split}
A +C
=\left(
\begin{array}{cc}
\frac{1}{1 + \lambda \lambda^T}  &\frac{1}{1 + \lambda \lambda^T}  \lambda \\
\frac{1}{1 + \lambda^T \lambda} \lambda^T  &  \frac{1}{1 + \lambda^T \lambda} \lambda^T \lambda
\end{array}
\right)  \equiv \tilde A \ .
\end{split}
\ee
The matrix $\tilde A$ 
can then be diagonalized by an action of the $O(N)$ group without affecting the bulk kinetic term, as $\tilde A$ is
symmetric with real coefficients. 
Note that $\tilde A ({\rm id} -\tilde A) = 0$, guaranteeing that all the eigenvalues of $\tilde A$ are either zero or one.  
%
We  ``diagonalize'' the theory with a change of basis matrix $\phi = B \cdot \varphi$, after which the action remains of the 
form (\ref{Sscalar}) but with $\lambda_{ai} = 0$:
\be
S_{\rm bry}  =  \int_{x^\perp=0} \d^{d-1} x \, \left( \varphi_a  \partial_\perp \varphi_a \right) \ .
\ee

We play one further game with the matrices $A$ and $\tilde A$.  We define a new matrix $\chi$ such that
\be
\tilde A = \frac{1}{2} ({\rm id} - \chi) \ .
\ee
As $\tilde A$ is a projector, it follows that $\chi^2 = {\rm id}$.  
This matrix will be useful in defining two-point functions of the $\phi$ fields later on.  In terms of $\chi$, the boundary conditions
(\ref{scalarbc}) can be written more compactly
\be
\label{scalarbcchi}
(1-\chi)\phi = 0 \ , \; \; \; (1+\chi) \partial_\perp \phi = 0 \ .
\ee

As a simple example, consider the $O(2)$ case where 
\be
\tilde A(\theta) =  \left( \begin{array}{cc} \cos^2 \theta & \cos \theta \sin \theta  \\ \cos \theta \sin \theta  & \sin^2 \theta  \end{array}\right) \ , 
\ee
\be
B =  \left( \begin{array}{cc} \cos \theta & -\sin \theta \\ \sin \theta & \cos \theta \end{array}\right)  \ ,
\ee
and $\lambda = \tan \theta$.  
Note the $\tilde A$ matrix has periodicity under $\theta \to \theta + \pi$ instead of $2\pi$ because the action (\ref{Sscalar}) 
has a ${\mathbb Z}_2$ symmetry and is invariant under $\phi_i \to -\phi_i$.
The theories $\tilde A(\theta)$ and $\tilde A(\theta+\pi/2)$ are related by permuting the $\phi_1$ and $\phi_2$ fields
and sending $\lambda \to -\frac{1}{\lambda}$, which is 
an example of a duality discussed by Witten \cite{Witten:2001ua} in an AdS/CFT context.

More generally, we have a situation where the boundary conditions 
break the $O(N)$ symmetry to $O(p) \times O(q)$ and the marginal couplings $\lambda_{ai}$ can be understood as 
coordinates on the coset moduli space $O(N) / (O(p) \times O(q))$, which is a real Grassmannian manifold.  
To characterize the moduli space using the field theory, we calculate the 
two, three and four point functions of the marginal operators ${\mathcal O}_{[ai]} = \phi_i \partial_\perp \phi_a$.  

The two point function of ${\mathcal O}_{[ai]}$, which gives the metric on the moduli space \cite{Zamolodchikov:1986gt},
follows from the two point function 
of the fundamental field $\phi_I$ by Wick's Theorem.  
This correlation function of $\phi_I$ with itself 
in turn is uniquely fixed by the condition that $\Box_x \langle \phi_I (x) \phi_J (y) \rangle = -\delta(x-y) \delta_{IJ}$
along with the boundary conditions.  The result is that
\be
\langle \phi_I(x) \phi_J(y) \rangle_\lambda = \kappa \left( \frac{\delta_{IJ}}{|x-y|^{d-2}} + \frac{\chi_{IJ}(\lambda)}{|\tilde x - y|^{d-2} }\right) \ .
\ee
We have introduced the normalization $\kappa^{-1} = (d-2) \Vol(S^{d-1})$ and the mirror point $\tilde x = (-x_\perp, x_\parallel)$.  This result can be reproduced by treating the moduli using conformal perturbation theory \cite{Sen:2017gfr,Herzog:2021spv}. In the boundary limit, we can then read off
\be
\begin{split}
\langle \phi(x_\parallel) \phi(0) \rangle_\lambda &= \kappa \frac{{\rm id} +\chi}{|x_\parallel|^{d-2}} \ , \\
\langle \partial_\perp \phi(x_\parallel) \partial_\perp \phi(0) \rangle_\lambda &= (d-2) \kappa \frac{{\rm id} -\chi}{|x_\parallel|^d} \ .
\end{split}
\ee
These correlation functions clearly satisfy the boundary conditions (\ref{scalarbcchi}) because $\tilde A$ is a projector.

Naively $\langle \phi (x_\parallel) \partial_\perp \phi (0) \rangle$ is zero as these two operators have different conformal dimension.
However, the full story is more complicated as these operators are also shadow dual.  By placing both insertions at
a height $\epsilon$ above the boundary and carefully taking the limit $\epsilon \to 0$, we find the contact term
\be
\label{contactterm}
\langle \phi(x_\parallel ) \partial_\perp \phi(0) \rangle = \frac{\chi}{2} \delta^{d-1}(x_\parallel) \ .
\ee

Applying Wick's Theorem, we find up to contact terms that we drop, that
\be
\langle {\mathcal O}_{[ai]} (x_\parallel) {\mathcal O}_{[bj]} (0) \rangle_\lambda = \frac{g_{[ai][bj]}(\lambda)}{|x_\parallel|^d} \ ,
\ee
where 
\be
\begin{split}
g_{[ai][bj]}(\lambda) &= \frac{L^2}{4} (\delta_{ab} - \chi_{ab} ) (\delta_{ij} + \chi_{ij}) \\
&= L^2 \left(\frac{1}{1+\lambda \lambda^T} \right)_{ab} \left( \frac{1}{1+\lambda^T \lambda} \right)_{ij} \ ,
\label{scalarmetric}
\end{split}
\ee
can be interpreted as a metric on the coset space $O(N) / (O(p) \times O(q))$, where we have defined the ``length scale''
$L^2 \equiv 4(d-2) \kappa^2$.  
A simple example to give us confidence this metric is correct is the case $p=1$ and $q=2$, where the moduli space is $S^2$.  By making the variable substitutions $\lambda_{1,2} = \tan \theta \cos \phi$ and $\lambda_{1,3} = \tan \theta \sin \phi$, we recover the usual round metric on an $S^2$: $\d s^2 = L^2 ( \d \theta^2 + \sin^2 \theta \d \phi^2)$. 
Indeed, (\ref{scalarmetric}) is the unique $O(N)$-invariant metric on the Grassmannian, up to the scale $L$ \cite{Leichtweiss:1961aa,Wong1967,Besse2007}.
(There is an exception for $O(4)$.  We will discuss the uniqueness result in more detail at the end.)
 

The $O(N)$ invariance of this metric is not obvious, except for the $O(p)\times O(q)$ subgroup. To find the transformation law of the $\lambda$ under the remaining transformations, note that we can identify a point in $O(N)/ (O(p) \times O(q))$ with the set $\{\left(\lambda v, -v\right)\in \mathbb{R}^{N}, \forall v\in \mathbb{R}^{q}\}$. This vector transforms in the usual vector representation. In this parametrisation, we can perform an infinitesimal transformation in the off-diagonal sector to identify the induced change in $\lambda$. This gives $\delta \lambda = \rho + \lambda \rho^{T}\lambda$, $\rho\in\mathbb{R}^{p\times q}$. This non-linearly realised $O(N)$ transformation law defines a Killing vector of this metric.
  
From the metric (\ref{scalarmetric}), it is straightforward to use the usual rules of Riemannian geometry to compute a connection and a Riemann curvature tensor.  By direct computation, we find
\begin{align}
\label{Gamma}
	\Gamma^{[ck]}_{[ai][bj]}(\lambda)&=\frac{1}{2}\left( \delta_{a}^{c}\chi_{bi}(\lambda) \delta_{j}^{k}+\delta_{b}^{c}\chi_{aj}(\lambda) \delta_{i}^{k}\right)  , \\
	R_{[c k][a i][d l][b j]}&= 
	 \frac{1}{L^2}\biggl( g_{[a i][c j]} g_{[b k][d l]} -g_{[a i][d k]}g_{[b j][c l]}
	  \nonumber \\
	 & + g_{[a i][b k]} g_{[c j][dl]}
	  - g_{[ai][c l]} g_{[b j][d k]} \biggr) \ .
	  \label{riemann}
\end{align}
We can also compute the Ricci tensor and scalar -- $R_{[ai][bj]} = \frac{N-2}{L^2} g_{[ai][bj]}$ and $R = \frac{N(N-2)}{L^2}$ -- indicating that the Grassmannian has been endowed with an Einstein metric with positive curvature. In the special case $p=1$ or $q=1$, 
the conformal manifold is the sphere $S^{N-1}$, and the Riemann tensor takes the form $R_{abcd} = \frac{1}{L^2} (g_{ac} g_{bd} - g_{ad} g_{bc})$.  

This direct computation is not always practical when studying conformal manifolds, as the correlators are usually known only in a small neighborhood near specific points, for example near $\lambda = 0$. It is instructive to compare these direct results with the usual operatorial approach, which involves computing integrated three and four point functions of the marginal operators ${\mathcal O}_{[ai]}$.



Marginal operators have a special OPE structure \cite{Seiberg:1988pf,Kutasov:1988xb}, with contact terms given by other moduli. Using the Wick contraction (\ref{contactterm}), we can see here explicitly that
\begin{align}
	\mathcal{O}_{A}(x_\parallel)\mathcal{O}_{B}(0)&=\frac{g_{AB}(\lambda)}{|x_\parallel|^{2d-2}}+\Gamma^{C}_{AB}
	\delta^{d-1}(x_\parallel)\mathcal{O}_{C}(0)+ \ldots
\end{align}
where we have introduced the multi-index $A = [ai]$.
The contact terms induce a mixing of the operators which correspond to the connection of the metric. This specific form of the OPE fixes the three point function, which only contains contact terms. This is necessary for a vanishing beta function. Otherwise, the power-laws would yield log divergences and a non-trivial scale dependence.

As a final exercise, we compute the integrated four-point function of the marginal operators ${\mathcal O}_{[ai]}$
and relate it to the Riemann curvature.   It is often challenging to perform the integrals, and various techniques have been
developed to simplify and streamline the procedure \cite{Friedan:2012hi,Balthazar:2022hzb}.  
Indeed, \cite{Drukker:2022pxk,Drukker:2022txy} have already employed the method of \cite{Friedan:2012hi} in computing
the curvature for line and surface defects.
Here however
we are able to proceed directly.
From the usual rules of quantum field theory, the second derivative of a two point function can be expressed as an integrated, connected, 
four point function:
\begin{align}
\lefteqn{	\partial_{[ai][bj]}g_{[ck][dl]}(\lambda)= }\\
& \int d^{d-1}x_1 d^{d-1}x_2 \langle \mathcal{O}_{[ai]}(x_1)\mathcal{O}_{[bj]}(x_2)\mathcal{O}_{[ck]}(x_0)\mathcal{O}_{[dl]}(0)\rangle_{\lambda} \nonumber
\end{align} 
with $x_0\cdot x_0 = 1$. At $\lambda = 0$, we can neglect the connection and the curvature can be expressed purely in terms of second derivatives of the metric
\begin{align}
\label{Rfromddg}
	R_{[ai][bj][ck][dl]}&=\frac{1}{2}\big(\partial_{[ck][ai]}g_{[dl][bj]}-\partial_{[ck][bj]}g_{[dl][ai]}\nonumber \\
	& -\partial_{[dl][ai]}g_{[ck][bj]}+\partial_{[dl][bj]}g_{[ck][ai]}\big) \ .
\end{align}
We would like to verify this formula. The four point function follows from Wick contractions. One should note 
this correlator contains contact terms. These contact terms are proportional to contractions $\langle \phi_i \partial_\perp \phi_a\rangle \sim \chi_{ia}  \propto \lambda$ that vanish at leading order. For this reason, we will ignore them. 
(One one can keep them to compute $\partial^2 g$ for generic $\lambda$.)

Let us evaluate this second derivative: 
\begin{eqnarray}
\label{secderiv}
\lefteqn{	\partial_{[ai][bj]}g_{[ck][dl]}(\lambda)= \kappa^4  \int
\frac{  d^{d-1}x_1 d^{d-1}x_2}{(\abs{x_1} \abs{x_{10}}\abs{x_2}\abs{x_{20}})^{d-2}} 
}
\\
&& \biggl(  \frac{ g_{[ai][dk]}g_{[bj][cl]}}{ \abs{x_{10}}^{2}\abs{x_2}^{2}} 
	+\frac{g_{[ai][cl]}g_{[bj][dk]}}{\abs{x_1}^{2} \abs{x_{20}}^{2}} 
	+ \frac{g_{[ai][dj]}g_{[bl][ck]}}{\abs{x_{12}}^{2}}  \nonumber \\
	&&+\frac{g_{[ai][bl]}g_{[ck][dj]}}{\abs{x_1}^{2}\abs{x_{20}}^{2}}+ \frac{g_{[ai][bk]}g_{[cj][dl]}}{\abs{x_2}^{2}\abs{x_{10}}^{2}}
	+\frac{g_{[ai][cj]}g_{[bk][dl]}}{\abs{x_{12}}^{2}}\biggl) \ . \nonumber
\end{eqnarray}
There are two types of terms:  power laws involving only $x_1$ and $x_2$, which take a factorised form;
and power laws involving $x_1$, $x_{20}$ and $x_{12}$ (or equivalently with $x_1$ and $x_2$ exchanged). We regularize these expressions by treating them as distributions and replacing them with their formal Fourier transform. We use the following two formulae:
\begin{align}
\begin{split}
	\frac{2 \kappa}{r^{d-2}}&= \int \frac{d^{d-1}k}{(2\pi)^{d-1}} \frac{e^{i k\cdot r}}{k}  \ , \\
	 \frac{2 \kappa }{r^{d}}&= -\frac{1}{d-2} \int \frac{d^{d-1}k}{(2\pi)^{d-1}} e^{i k\cdot r}k \ , 
	 \end{split}
\end{align}
from which we are able to perform the position space integrals. We find through this prescription that we can set the first two terms of (\ref{secderiv}) to zero.  The relevant integrals are of the form 
\begin{align}
	\int d^{d-1}x_1\frac{1}{\abs{x_1}^{d-2}\abs{x_{10}}^{d}}	 \propto \delta^{d-1}(x_0) = 0 \ .
\end{align}
For the other terms, we use
\begin{align}
	\int \frac{d^{d-1}x_1d^{d-1}x_2}{\abs{x_1}^{2 \Delta+2}\abs{x_{12}}^{2\Delta}\abs{x_{20}}^{2 \Delta+2}} =-\frac{1}{L^2} \ .
\end{align}

The end result of these manipulations is that we fix from the operatorial prescription 
\be
\lefteqn{
\partial_{[ai][bj]}g_{[ck][dl]}= -\frac{1}{L^2}\biggl( g_{[ai][dj]}g_{[bl][ck]}+g_{[ai][bl]}g_{[ck][dj]} }
\nonumber \\
	&& \hspace{0.8in} +g_{[ai][bk]}g_{[cj][dl]}+g_{[ai][cj]}g_{[bk][dl]}\biggl) 
\ee
at the point $\lambda = 0$.  
With this explicit result, we can compute the Riemann tensor from (\ref{Rfromddg}). We find back the result (\ref{riemann}) previously derived, evaluated at $\lambda=0$.

\vskip0.1in

\noindent
{\it Spinor Moduli}

The starting point for the spinors is a system of $N$ free massless spinor fields $\psi_I$, $I = 1, \ldots, N$, again in the presence of a boundary at $x_\perp = 0$.  The existence of the $\gamma^\perp$ gamma matrix allows us to define a pair of projectors
$P_\pm = \frac{1}{2}(1 \pm \gamma^\perp)$.  The initial boundary conditions consist of imposing $P_+ \psi_a = 0$ on $p$ of the spinors and $P_- \psi_i = 0$ on $q$ of the spinors, where as before $p+q = N$.  The global $U(N)$ symmetry group is then broken to $U(p) \times U(q)$. For our conventions, see \cite{Herzog:2022jlx}. 
%
%

The marginal operators are again the boundary limit of the normal component of the conserved currents, 
$J^\mu = \bar \psi_I \gamma^\perp \psi_J$.  To discuss what happens to the boundary limit of $J^\perp$ having
applied the above boundary conditions, it is useful to introduce the projected boundary values of the spinor
\be
\rho_I^\pm = P_\pm \psi_I \ .
\ee
Because of the presence of $\gamma_0$ in the definition of $\bar \psi$, only the combinations  
$\bar \rho_I^\pm \gamma^\perp \rho_J^\mp$ survive in the boundary limit of $J^\perp = \bar \psi_I \gamma^\perp \psi_J$.

The action that describes our system, including the deformation by the marginal operators, is 
\begin{align}
\begin{split}
S &= - \frac{1}{2}\int_{x^{\perp}>0} d^{d}x \left( \overline{\psi}_I \slashed{\partial}\psi_I -\partial_\mu\overline{\psi}_I\gamma^{\mu}\psi_I \right)  \\
&\hspace{0.2in} + \frac{1}{2} \int_{x^\perp = 0}  d^{d-1}x \,  \left( \overline{\rho}^+ A^\dagger \rho^- + \overline{\rho}^- A \rho^+ \right) \ .
\end{split}
\end{align}
where
\begin{align}
	A &= \begin{pmatrix}
		{\rm id}_p & 2\lambda\\ 0 & -{\rm id}_q
	\end{pmatrix} \ , 
\end{align}
Like in the bosonic case, this choice for $A$ imposes boundary conditions on the field which take the form 
\begin{align}
	{\lambda}^\dagger_{i,a}\rho^{-}_{a} &= +\rho^{-}_{i} \ , & \lambda_{a,i}\rho^{+}_{i} &= -\rho^{+}_{a} \ . 
\end{align}

Again, one can add the redundant operator $\bar \rho^+ C \rho^-$ to the action to modify $A$, where
\begin{align}
C \equiv 
 \left( \begin{array}{cc}
-\frac{1}{1 + \lambda \lambda^\dagger} \lambda \lambda^\dagger &-\frac{1}{1 + \lambda^\dagger \lambda} \lambda^\dagger \lambda \lambda^\dagger \\
\frac{1}{1 + \lambda \lambda^\dagger}  \lambda & \frac{1}{1 + \lambda^\dagger \lambda} \lambda^\dagger \lambda
\end{array}
\right) 
\end{align}
and the Hermitian conjugate to modify $A^\dagger$. The resulting 
\be
\tilde A \equiv A  + C 
	=  \begin{pmatrix}
		-{\rm id}+\frac{2}{1+{\lambda^\dagger}\lambda} & \frac{2}{1+{\lambda^\dagger}\lambda} \lambda^\dagger \\ \frac{2}{1+\lambda{\lambda^\dagger}}\lambda& {\rm id} -\frac{2}{1+\lambda {\lambda^\dagger}}{}
	\end{pmatrix} \ 
\ee
is Hermitian and can be diagonalized by a $U(N)$ transformation, i.e.\ an element of the global
symmetry group of the bulk theory.  Thus the deformation by $\lambda$ can be undone by a field redefinition, just like in the scalar case.
The big difference from the scalar case
 is that the $\lambda$ are complex and are coordinates on the complex Grassmannian manifold, the coset
$U(N) / (U(p) \times U(q))$.  
To keep the notation parallel to the scalar case, we introduce $\chi \equiv \tilde{A}$.  

The bulk $\psi_I$ two-point function is fixed by boundary conditions and the equation of motion to be
\begin{align}
	\expval{\psi_I(x)\overline{\psi}_{J}(y)}&= \kappa_f \left( \delta_{IJ}\frac{\gamma\cdot(x-y)}{\abs{x-y}^{d}}+\chi_{IJ}\frac{\gamma_\perp\gamma\cdot(\tilde{x}-y)}{\abs{\tilde{x}-y}^{d}}\right) \ , 
\end{align}
where $\kappa_f \equiv (d-2) \kappa$. Again, this result can be reproduced independently using perturbation theory.  The boundary limits of this two-point function obey
\begin{align}
	\expval{\rho^{\alpha}_{I}(x)\overline{\rho}^{\alpha}_{J}(0)}&=\kappa_f(\delta_{IJ}+\alpha \chi_{IJ})\frac{\gamma\cdot x P_{-\alpha}}{\abs{x}^{d}} \ ,  \\
	\expval{\rho^{\alpha}_{I}(x)\overline{\rho}^{-\alpha}_{J}(0)}&
	= \frac{\chi_{IJ}}{2}P_{\alpha}\delta^{d-1}(x) \ .
\end{align}

The next step is to compute the metric on the conformal moduli space for general values of $\lambda$.
The metric can be computed from Wick's Theorem applied to the
marginal operators $\mathcal{O}_{[ai]}= \overline{\rho}^{+}_{a}\rho^{-}_{i}$ and complex conjugate:
\be
\expval{\mathcal{O}_{[ai]}(1)\overline{\mathcal{O}}_{\overline{[jb]}}(0)}_\lambda = g_{[ai]\overline{[jb]}}(\lambda) \ .
\ee
The metric is evidently for a complex manifold.
We find 
\begin{align}
	 g_{[ai]\overline{[ja]}}(\lambda)&=L^2 \left(\frac{1}{1+\lambda \lambda^\dagger}\right)_{ab}\left(\frac{1}{1+\lambda^\dagger \lambda}\right)_{ij}
\end{align}
where $L^2 = 2 \kappa_f^2 D$ and $D$ is the dimension of the spinor representation.  The metric 
is the standard $U(N)$ invariant Fubini-Study type metric on the complex Grassmannian $U(N) / (U(p) \times U(q))$. 
It is again unique \cite{Leichtweiss:1961aa,Wong1967,Besse2007}.
Unlike the real case, It can be derived from a K\"ahler potential: 
\begin{align}
	g_{[ai]\overline{[jb]}} &= \pdv{}{\lambda_{[ai]}}\pdv{}{\lambda^\dagger_{\overline{[jb]}}}K 
\end{align}
where $K = L^2 \log \det (1+\lambda {\lambda^\dagger})$.

The Ricci tensor follows from the standard formula
\begin{align}
	R_{[ai][bj]}&= -\pdv{}{\lambda_{[ai]}}\pdv{}{{\lambda^\dagger}_{\overline{[jb]}}}\log \det g = \frac{N}{L^2}g_{[ai]\overline{[jb]}}
\end{align}
indicating the manifold is Einstein with positive curvature.  The calculation requires the Weinstein-Aronszajn identity, that
$\det ({\rm id} + \lambda^\dagger \lambda) = \det ({\rm id} + \lambda \lambda^\dagger)$.  
The Ricci scalar finally is $R  = N^2/ L^2$.  

\vskip 0.1in

\noindent
{\it Discussion}

This work explained how the interplay between continuous symmetry and boundary conditions gives rise to fully computable examples of conformal manifolds.  
While we have performed computations in two free theories, the free scalar and spinor, we anticipate that this conformal manifold produced by inserting tilt operators into the action will be insensitive to the presence of interactions that preserve the bulk global symmetry -- $O(N)$ in the scalar case, $U(N)$ in the spinor case.  In each case the conformal manifold was a Grassmannian, and in each case the Grassmannian was endowed with essentially a unique metric invariant under global symmetry.  Thus the interactions can do nothing to perturb the metric, except change the scale $L$.  
Because of the bulk group action, all points on the manifold should look the same.  In other words, the manifold is homogeneous.  
One has the following theorem due to Wolf \cite{Wolf3}: Let $M = G/K$ be a $G$-homogeneous isotropy irreducible space.  Then (up to homotheties) $M$ admits a unique $G$-invariant Riemannian metric.  This Riemannian metric is Einstein.  

For our $O(N) / (O(p) \times O(q))$ and $U(N) / (U(p) \times U(q))$ cosets, the one exception to the isotropy condition occurs
for $O(4) / (O(2) \times O(2))$ because $O(4)$ breaks apart into a product of $O(3)$ groups.  For the corresponding metric on a product of two-spheres, the radii of the spheres can be adjusted independently.  It would be interesting to see if such a situation can be realized through an interacting $O(4)$ model of scalar fields.  In all the other cases that we studied, the metric is unique up to rescaling (homothety).  

A richer class of conformal manifolds will emerge from field theories 
that support different classes of boundary conditions \footnote{%
 See already 
 \cite{Drukker:2022txy} 
 for a coset involving a product of three quotient groups in the line defect setting.
 }.
 One candidate 
are higher derivative theories \cite{Chalabi:2022qit}.  A theory with an action $\int \d^d x \phi \Box^2 \phi$ for instance
supports four different types of boundary conditions, allowing a  breaking pattern 
$O(N) / (O(p) \times O(q) \times O(r) \times O(N-p-q-r))$. 
We leave investigations of higher derivative and supersymmetric
bCFTs to future work.

%

\vskip 0.1in
\noindent
{\it Acknowledgments:}
We would like to thank J\"urgen Berndt, Hans Werner Diehl, Nadav Drukker, George Sakkas, and Andy Stergiou for discussion.
This work was funded in part by  a Wolfson Fellowship from the Royal Society
  and by the U.K.\ Science \& Technology Facilities Council Grant ST/P000258/1.

\bibliography{moduli}

\end{document}